# Replica symmetry breaking in 1D Rayleigh scattering system: theory and validations

Yifei Qi[1], Longqun Ni[1], Zhenyu Ye[1], Jiaojiao Zhang[1], Xingyu Bao[1], Pan Wang[1], Yunjiang Rao[1], Ernesto P. Raposo[2,*], Anderson S. L. Gomes[3,*] and Zinan Wang[1,*]

[1] Key Lab of Optical Fiber Sensing & Communications, University of Electronic Science and Technology of China (UESTC), Chengdu, China

[2] Laboratório de Física Teórica e Computacional, Departamento de Física, Universidade Federal de Pernambuco, 50670-901 Recife, Pernambuco, Brazil

[3] Departamento de F́ısica, Universidade Federal de Pernambuco, Recife, Pernambuco, Brazil

* ernesto.raposo@ufpe.br

* andersonslgomes@gmail.com

*znwang@uestc.edu.cn

## Abstract

Spin glass theory, as a paradigm for describing disordered magnetic systems, constitutes a prominent subject of study within statistical physics. Replica symmetry breaking (RSB), as one of the pivotal concepts for the understanding of spin glass theory, means that, under identical conditions disordered systems can yield distinct states with nontrivial correlations. Random fiber laser (RFL) based on Rayleigh scattering (RS) is a complex disordered system, owing to the disorder and stochasticity of RS. In this work, for the first time, we elaborate a precise theoretical model for studying the photonic phase transition via the platform of RS-based RFL, in which we clearly reveal that, apart from the pump power, the photon phase variation in RFL is also an analogy to the temperature term in spin glass phase transition, leading to a novel insight into the intrinsic mechanisms of photonic phase transition. In addition, based on this model and real-time high-fidelity detection spectral evolution, we theoretically predict and experimentally observe the mode-asymmetric characteristics of photonic phase transition in RS-based RFL. This finding contributes to a deeper understanding of the photonic RSB regime and the dynamics of RS-based RFL.

## Introduction

The spin glass theory gained widespread attention since the 1970s, serving as a framework to describe a critical state in disordered magnetic systems in which nontrivially correlated spins

"freeze" at random directions below some critical temperature [1-4]. Subsequently, Sherrington and Kirkpatrick made significant contributions by proposing a mathematical model based on Ising model, where spins are coupled by an infinite range of random interactions [5], thereby conducting an in-depth exploration of the spin glass theory. To provide a more comprehensive description of complex phenomena within spin glass systems and circumvent technical difficulties related to the lack of stability of the Sherrington and Kirkpatrick solution, Parisi introduced the RSB theory [6,7]. When a system is in a spin glass state, it exhibits numerous local minima associated with configuration states in the free energy landscape. In this regime, identical replicas under the same experimental condition can manifest distinct properties, a phenomenon known as RSB, effectively characterized using the Parisi overlap parameter [8]. Considering the importance of RSB not only for statistical physics, but also its connection to turbulence behavior [9], the Nobel Prize in Physics in 2021 was partially laureated to Giorgio Parisi "for the discovery of the interplay of disorder and fluctuations in physical systems from atomic to planetary scales". Research into the spin glass theory contributes to the understanding of the properties of complex systems characterized by non-uniform and disordered attributes, and extends its applicability to fields such as materials science, neural networks, and condensed matter physics [10-13].

Recently, the RSB spin-glass theory has been applied to investigate various disordered systems, including random lasers and nonlinear wave propagation [14-16]. In 2015, Ghofraniha et al. made a significant breakthrough by observing the RSB phenomenon for the first time in experiments using a solid-state random laser platform. They inferred the Parisi overlap parameter from fluctuations in spectral intensity and the experiments revealed that, as the pump power transitions from below the threshold to above the threshold, the random laser undergoes a phase transition from a photonic paramagnetic state to a spin glass state [17]. In 2016, Gomes et al. demonstrated simultaneous observation of RSB and Lévy behavior in a Nd-powder cased RL[18], and Pincheira et al. reported the occurrence of RSB in a colloidal based RL[19]. In the same year, Tommasi et al. conducted experiments confirming the RSB phase transition phenomenon in different disordered random lasers [20]. In 2017, Pierangeli et al. observed the RSB phenomenon in the context of nonlinear wave propagation [21].

The experiments mentioned above have predominantly utilized complex three-dimensional waveguide materials, which are sensitive to environmental factors and present challenges in generating replicas. In contrast, RFL provides a one-dimensional platform for studying spin glass theory, which encapsulates disorder in feedback and gain within a one-dimensional optical fiber waveguide [22], and a review on the work on RFL related to the Physics Nobel Prize in 2021 is reported [23]. The core idea for investigating the spin glass theory in photonic systems, such as RLs and RFLs, is the analogy of the lasing modes to spin variables [24]. In addition, these systems, akin to an inverse-temperature system, exhibit different laser states under varying pump energies, transitioning from a photonic paramagnetic phase at low pump energies to a spin glass phase at high pump energies. Gomes et al. first observed RSB in erbium-doped fiber random lasers, introducing disorder through a random fiber grating [25,26]. While RSB has been observed in diverse disordered systems, these experiments tend to require long time intervals for replica acquisition and lack a corresponding numerical simulation platform.

In this work, the intrinsic mechanisms of photonic phase transition in a RS-based RFL system are investigated through both theoretical and experimental approaches, for the first time, to the best of our knowledge. Compared to random fiber grating, RS has a higher degree of disorder. Benefiting from the precise calibration of RS phase fluctuations over time with unprecedented precision, a RS-phase-variation model of RS-based RFL is proposed, which is well validated by high-precision spectral measurements (1 $\mu$s time-domain resolution and 2 MHz frequency-domain resolution). Then, based on this model and real-time detection of high-fidelity spectral evolution, we find not only that the photon phase variation in RS-based RFL can be analogous to the temperature in the phase transition, but also that the photonic phase transition in RS-based RFL exhibits mode-asymmetry. This finding not only provides a novel experimental and simulation platform for RSB research but also advances our understanding of the complex systems.

## Theoretical analysis

As a start, we performed corresponding numerical analysis to theoretically study the dynamic characteristics of the RS-based RFL. The simulation model is based on the generalized nonlinear Schrödinger equations (NLSEs) [27,28]:

$$\frac{\partial u_p^\pm}{\partial z} \mp \frac{1}{v_{gs}}\frac{\partial u_p^\pm}{\partial t} \pm i\frac{\beta_{2p}}{2}\frac{\partial^2 u_p^\pm}{\partial t^2} \pm \frac{\alpha_p}{2}u_p^\pm = \pm i\gamma_p|u_p^\pm|^2 \mp \frac{g_p(\omega)}{2}\left(\langle|u_s^\pm|^2\rangle + \langle|u_s^\mp|^2\rangle\right)u_p^\pm, \qquad (1)$$

$$\frac{\partial u_s^\pm}{\partial z} \pm i\frac{\beta_{2s}}{2}\frac{\partial^2 u_s^\pm}{\partial t^2} \pm \frac{\alpha_s}{2}u_s^\pm \mp \frac{\varepsilon(\omega)}{2}u_s^\pm = \pm i\gamma_s|u_s^\pm|^2 u_s^\pm \pm \frac{g_s(\omega)}{2}\left(\langle|u_p^\pm|^2\rangle + \langle|u_p^\mp|^2\rangle\right)u_s^\pm, \quad (2)$$

where subindexes $'p'$ and $'s'$ represent the pump and Stokes waves respectively; $'+'$ and $'-'$ correspond to forward and backward light; $u$ is the envelope of the optical field; $v_{gs}$ is the group velocity difference arising from the wavelength discrepancy between the pump and Stokes waves; $\omega$ is the angular frequency of lightwave; $\alpha, \gamma, \beta_2, \varepsilon$ and $g$ are the linear fiber loss, Kerr coefficient, second-order dispersion, RS and Raman gain respectively, and $\varepsilon$ and $g$ are related to frequency in this simulation.

The boundary conditions can be described as:

$$P_p^+(0,\omega,t) = P_{in}(\omega)T_{L_p} + R_{L_p}(\omega)P_p^-(0,\omega,t), \ P_p^-(L,\omega,t) = R_{R_p}(\omega)P_p^+(L,\omega,t), \quad (3)$$

$$P_s^+(0,\omega,t) = R_{L_s}(\omega)P_s^-(0,\omega,t), \ P_s^-(L,\omega,t) = R_{R_s}(\omega)P_s^-(L,\omega,t), \quad (4)$$

where $R_L$ and $R_R$ are the reflection spectrum of the FBG and the fiber respectively; $P_{in}$ denotes the input pump power; $L$ is fiber length which is 15 km. The parameter values set in this simulation are shown in Table 1.

Table1 Parameters set in the simulation (see text for definitions).

| Parameter | Pump | Stokes |
| --- | --- | --- |
| $\lambda(nm)$ | 1455 | 1550 |
| $v_g(m/s)$ | $2.0504 \times 10^8$ | $2.0497 \times 10^8$ |
| $\alpha(dB/km)$ | 0.24 | 0.2 |
| $g(m^{-1}W^{-1})$ | $4.14 \times 10^{-4}$ | — |
| $\gamma(m^{-1}W^{-1})$ | 0.0017 | 0.0014 |
| $\varepsilon(m^{-1})$ | $0.6 \times 10^{-7}$ | $0.45 \times 10^{-7}$ |
| $\beta_2(s^2/m)$ | $-1.7324 \times 10^{-26}$ | $-2.7927 \times 10^{-26}$ |
| $R_L$ | $4 \times 10^{-5}$ | 0.99 |
| $R_R$ | $4 \times 10^{-5}$ | $4 \times 10^{-5}$ |

It should be noted that, as a pivotal parameter in RS-based RFL, the RS phase does not remain constant, but changes according to the external environment, and the degree of RS phase fluctuation will determine the excited state of RS-based RFL. Thus, the RS-based RFL with different fluctuating states of RS phase is simulated. As showing in Fig.1(a), the fluctuation of RS phase changed from 0.1 rad to 4.8 rad, and the corresponding RS-based RFL output characteristics are shown in Fig.1(b). From the simulation results we can get that: when the fluctuation of RS phase is small, such as 0.1 rad to 3.6 rad, the spectrum of the RFL exhibits randomly distributed spikes, and the number of the spikes decreases as the degree of RS-phase fluctuation increases; conversely, when the fluctuation of RS phase is large, such as 4.8 rad, the RFL has a smooth spectrum. The reason for this is as follows. The RS reflection spectrum has random distributed spikes [29], and when the fluctuation of the RS phase is small, the RS reflection spectrum can keep stable. Therefore, a stable coherent resonant cavity can be formed and random longitudinal modes are excited, which are manifested in the spectrum as randomly distributed spikes. However, when there is an external action causing large fluctuations in the RS phase, the reflection spectrum of the RS undergoes a dramatic change. Consequently, it fails to establish stable feedback, indicating that the random longitudinal modes cannot be established.

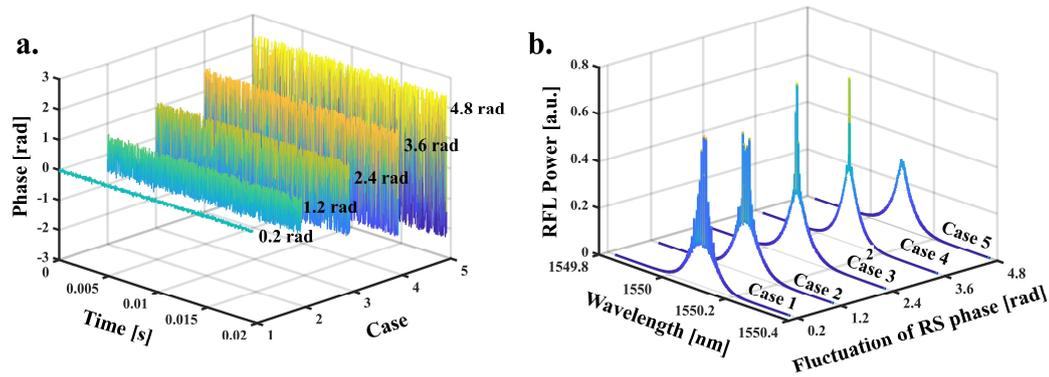

**Figure 1** (a) Five different fluctuation ranges of RS phase in simulation: Case 1: 0.2 rad, Case 2: 1.2 rad, Case 3: 2.4 rad, Case 4: 3.6 rad, Case 5: 4.8 rad. (b) Corresponding evolution of spectrum: as the degree of RS phase fluctuation increases, the randomly distributed spikes of RFL spectrum gradually disappear.

The above results show that, the degree of the RS phase fluctuation determines the excited state of RS-based RFL. Therefore, determining the degree of the fluctuations of the RS phase in the experimental environment is particularly important for accurately simulating the excitation state of the RS-based RFL, and it was measured by a $\phi$-OTDR with proprietary technologies [30]. This refinement process has contributed to the attainment of a high-fidelity representation of RFL behavior within the simulations.

# Experimental setup

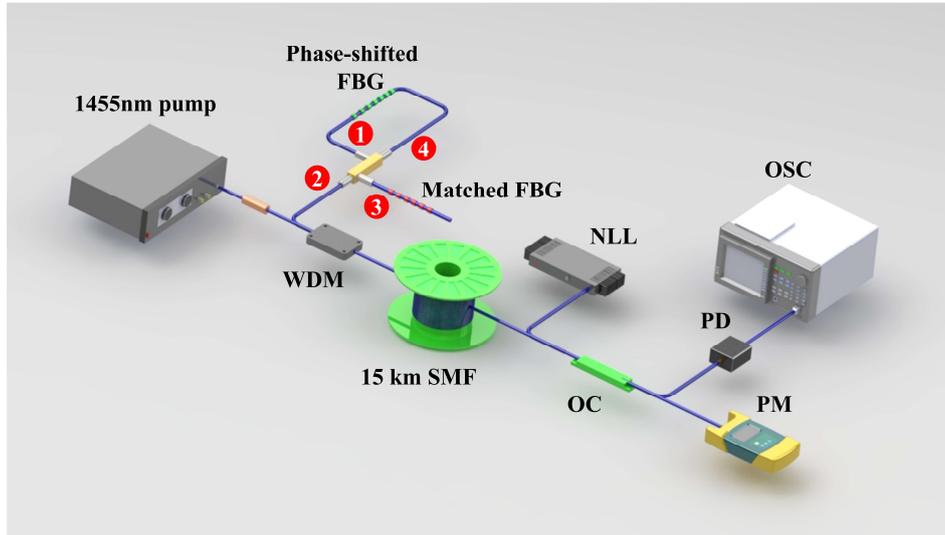

**Figure 2** Experiment setup for rapid spectrum detection. WDM: wavelength division multiplexer; SMF: single model fiber; OC: optical coupler; PD: photodetector; OSC: oscilloscope; PM power meter.

The setup of the experiment is shown in Fig.2, which is consistent with the simulation. The 1455 nm pump source injects into a 15 km single mode fiber (SMF) through a 1455/1550 nm wavelength division multiplexer (WDM) and the 1550 nm port is connected to the feedback. It is worthy to note that limited by the bandwidth of photodetector (PD), the transmission peak of the phase-shift FBG is chosen as the feedback of the RFL, which is 0.02 nm. The generated RFL outputs at the end of the fiber and in order to achieve rapid and high-precision spectral detection, a narrow linewidth tunable laser (NLL) was used in conjunction with the generated RFL for beat-frequency operation in the experiment [31,32], and the power meter is employed to monitor the laser power, preventing potential damage to the detector caused by excessive power. Time-domain data was collected using a 40 GHz bandwidth detector and a 16 GHz bandwidth oscilloscope, and the data acquisition duration for each dataset is 5ms.

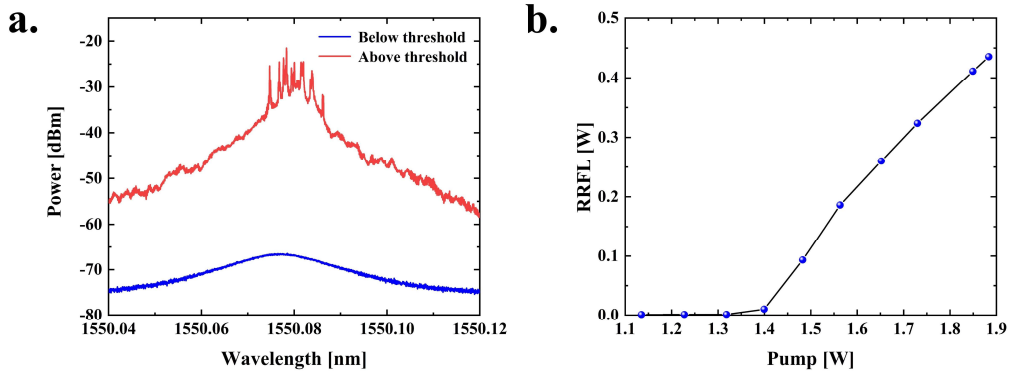

**Figure 3. a.** The spectrum of RFL for different pump power. When the pump power is above the threshold, randomly distributed spikes appear in the RFL spectrum. **b.** Output power of the RFL versus pump power

The spectrum of RFL is shown in Fig.3a, and Fig.3b shows the output RFL at the fiber end as a function of the pump power. The spectrum was measured by a high-resolution spectrometer with a resolution of 0.16 pm. When the pump power is below the threshold (which is 1.3 W), it is primarily spontaneous Raman scattering; when the pump power above the threshold, due to the stochasticity and incoherence of the RS, the spectrum of the RFL exhibits characteristics of randomly distributed peaks.

## Results and discussions

The relationship between the RS and output characteristics of RS-based RFL has been analyzed theoretically based on the RS-phase-variation model above. In order to further confirm our conclusions, we designed two validation experiments with different degree of RS phase fluctuation. In Case 1, the optical fiber of the RS-based RFL is in a stable state, in which the degree of fluctuation of the RS phase is small, and the spectrum of the RFL is shown as the blue curve in Fig. 4; otherwise, in Case 2, a significant external action is applied to the optical fiber of the RS-based RFL, in which the optical range between two points of RS will be changed and leading to a larger degree of RS phase fluctuation, and the spectrum is shown as the red curve in Fig. 4. Experimentally and theoretically, it is proved that when the RS phase fluctuation degree is small, a stable coherent feedback cavity can be formed in the fiber, and the random longitudinal mode modes are excited in the RFL (which are manifested as randomly distributed spikes on the spectrum); and when the RS phase fluctuation degree is large, the randomly distributed spikes of the spectrum disappear.

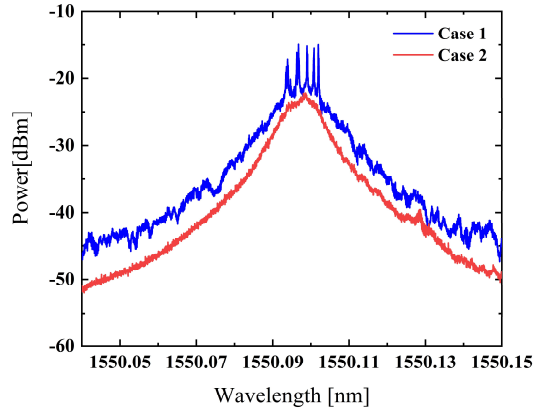

**Figure 4** The experimental RS-based RFL spectrum of different RS phase fluctuation: Case 1: small fluctuation of RS phase with the random spikes; Case 2: large fluctuation of RS phase without the random spikes.

To meet the bandwidth requirements, the frequency difference between the NLL and the RFL was set to 9 GHz. Figure 5a shows the simulated temporal spectrogram and Fig.5b is experimental result correspondingly. From the temporal spectrogram, we observe that the intensity of randomly distributed spikes on the spectrum are not static, but fluctuate over time. The evolution of the spectrum within a 5$ms$ interval is illustrated and the temporal resolution and wavelength resolution are 1$\mu s$ and 2$MHz$, respectively.

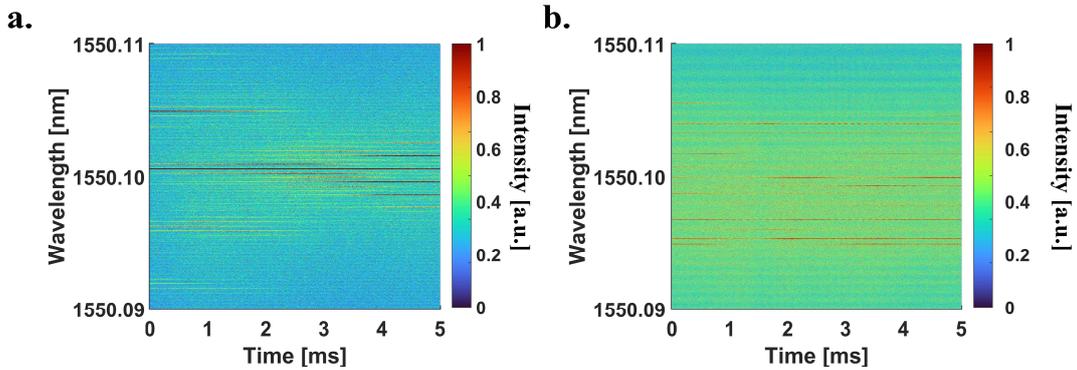

**Figure 5** Temporal spectrogram of RFL with randomly distributed spikes. **a.** Simulation. **b.** Experiment

The correlation among the RS-based RFL's wavelength can be explored through the Pearson's correlation coefficient involving the statistical covariance measure between intensity fluctuations at wavelengths $(\lambda_1, \lambda_2)$ in the same spectrum,

$$\rho(\lambda_1, \lambda_2) = cov(I_{\lambda_1}, I_{\lambda_2})/(\sigma_{\lambda_1}, \sigma_{\lambda_1}).$$

Figure 6 shows the matrix of $\rho$ values. The left-hand side shows the simulation results and the right-hand side shows the corresponding experimental results. Figure 6(a) and 6(b) illustrate the matrix of $\rho$ values below the threshold; Fig.6(c) and 6(d) present the matrix of $\rho$ values above the threshold with stable photon phase; Fig.6(e) and 6(f) present the matrix of $\rho$ values above the threshold with large fluctuation of photon phase. When the pump power is below the threshold, there is no mutual correlation among the spectral wavelengths. However, when the pump power exceeds the threshold, the spectral correlation arises between the randomly distributed spikes, while the remaining components of the spectrum remain uncorrelated. Due to the mutual interaction among the randomly distributed spikes, which shows a relationship of gain competition and gain sharing, the intensity of the narrowband components exhibits fluctuations. As shown in Fig.6(e) and 6(f), violent fluctuations of photon phase will disrupt the interactions between different longitudinal modes.

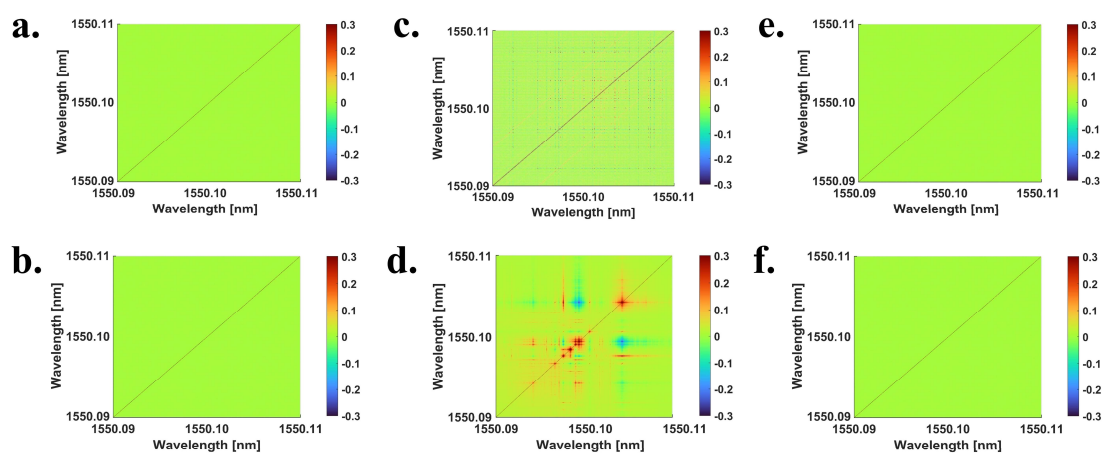

**Figure 6** Matrix of $\rho$ values: **a. b.** Below the threshold. The different RFL modes in the frequency domain are mutually independent, $\rho = 0$; **c. d. e. f.** Above the threshold. **c. d.** the randomly distributed spikes in the RFL spectrum exhibit correlation, while the remain spectral components are mutually independent; **e. f.** the photon phase variation will compromise the correlation between the random longitudinal modes; **a. c.** Simulation. **e. d.** Experiment.

In order to investigate the RSB phase transition based on spectral intensity fluctuations, we selected spectral data below and above the threshold separately. For the experiment, 1000 spectra constitute a set of replicas. It is worth to note that each dataset has a temporal span of 5*ms*, within which the experimental conditions remain consistent. So, each dataset can be regarded as a set of replicas. For the simulation, we also choose 1000 spectra as a set of replicas which are under the same simulation conditions. The intensity fluctuations within each set of replicates are as follows:

$$\Delta_\alpha(k) = I_\alpha(k) - \bar{I}(k)$$

where the $\bar{I}(k)$ is the average over replicas of each wavelength intensity. The overlap parameter of the replicates is defined [16] as

$$q_{\alpha\beta} = \frac{\sum_{k=1}^{N} \Delta_\alpha(k)\Delta_\beta(k)}{\sqrt{\sum_{k=1}^{N}\Delta_\alpha^2(k)}\sqrt{\sum_{k=1}^{N}\Delta_\alpha^2(k)}}.$$

Based on the cross-correlation of the spectrum, the set of values of $q_{\alpha\beta}$ is calculated and the probability distribution P(q) of $q_{\alpha\beta}$ values is determined for several pump powers.

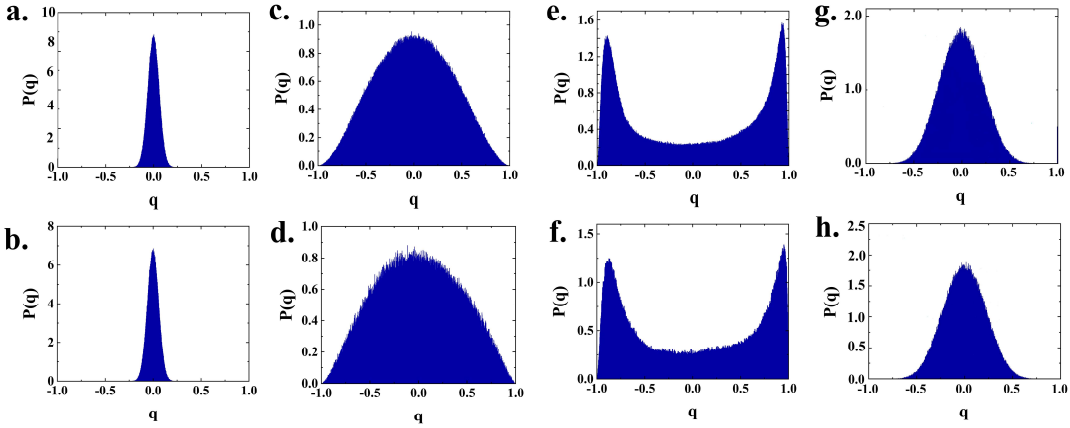

**Figure 7 Distribution function of the overlap q at different pump powers. a. b:** below the threshold, which means RFL is in paramagnetic state. **c. d. e. f. g. h.** above the threshold: **c. d.** randomly distributed spikes of RS-based RFL, which means the RSB has set in; **e. f.** remaining spectral components (the slowing varying spectral profile over wavelength and time) of RFL, which mean the remain spectral components is still in paramagnetic state; **g. h.** the photon phase variation disrupts the spin-glass state. Different parts of the spectrum exhibit different states consistent with their corresponding spectral correlations, which shows mode-asymmetric, and photon phase variation can be analogized to "high temperature". **a. c. e. g.** Simulation. **b. d. f. h.** Experiment.

Figure 7 shows the P(q) for different pump powers. The left-hand side shows the simulation results and the right-hand side shows the corresponding experimental results. As shown in Fig. 7(a) and 7(b), when pump power is below threshold, P(q) is centered around $q = 0$. In this case, we identify the maximum of the distribution P(q) as $q_{max} = 0$, which means that each mode of the RRFL is independent and essentially do not interact with others at the photonic paramagnetic phase. Remarkably, when the pump power exceeds the threshold, different components of the RFL exhibit entirely distinct characteristics, which is a surprising new discovery. Specifically, when the pump power surpasses the threshold, the RFL emits random longitudinal modes based on coherent feedback, appearing as randomly distributed peaks in the spectrum. Due to gain competition and gain sharing, these random longitudinal mode patterns are no longer independent; instead, they exhibit strong mutual relationships in the frequency domain. By extracting the randomly distributed peaks in the RFL spectrum as a set of replicas, the values of q are calculated. As illustrated in Fig.7(c) and 7(d), we find $q_{max} = 1$ (in absolute value), indicating that RSB has set in. However, the remaining components of the spectrum remains in the paramagnetic state, maintaining mutual independence in the frequency domain. By selecting the same data points from the remaining part of the spectrum as a set of replicas, now we find $q_{max} = 0$, as shown in Figures 7(e) and 7(f). Experimental and simulation results demonstrate the mode-asymmetry in the phase transition of the RFL. Additionally, as illustrated in the Fig. 7(g) and 7(h), when the photon phase of the RFL undergoes an intense fluctuation, the disappearance of the interactions among the random longitudinal modes leads to the overall alignment of the RFL in a paramagnetic state, akin to conditions analogous to elevated temperatures.

The relationship between $q_{max}$ and the pump power is depicted in Fig.8, and the experimental findings align with those obtained through simulation, exhibiting a congruent trend. As the pump power transitions from below the threshold to above it, $q_{max}$ changed from 0 to 1 which means a phase transition.

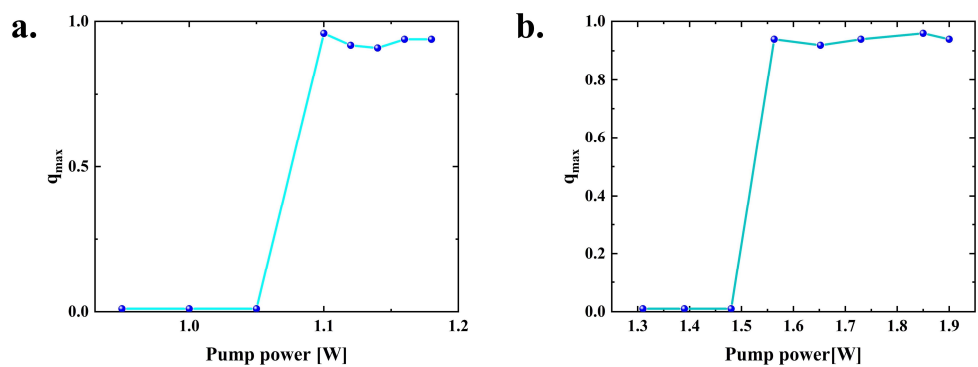

**Figure 8** The $q_{max}$ for random longitudinal modes of RFL versus pump power, which shows a phase transition as pump power increases. **a.** Simulation. **b.** Experiment.

Furthermore, the temporal correlation within narrowband components of the RFL was also investigated. Figure 9 shows the matrix of $\rho$ values for the temporal correlation. The left-hand side shows the simulation results and the right-hand side shows the corresponding experimental results. As shown in Fig.9, the narrowband components of the laser emission spectrum exhibit a high degree of temporal correlation. Through the manipulation of spectral randomly distributed spikes, such as the introduction of frequency shifts within the optical cavity, the correlation characteristics of the RFL can be controlled. The experimental results match well with the simulation results.

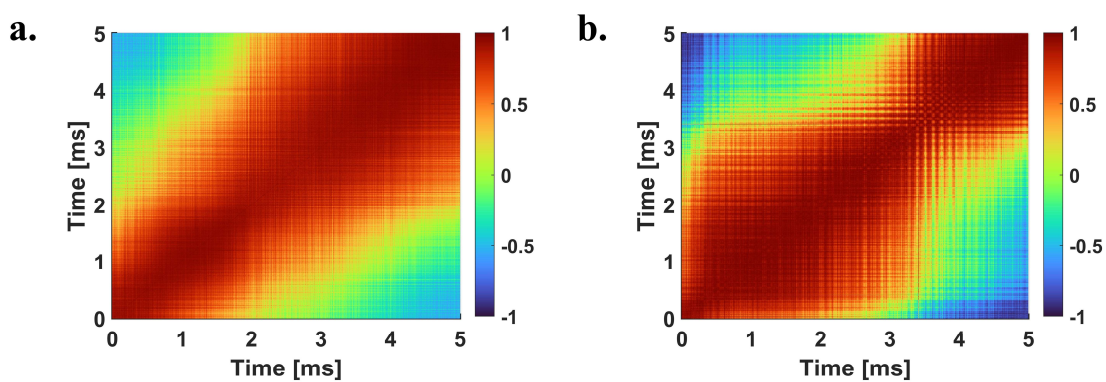

**Figure 9** Temporal correlation coefficient of the random distributed spikes of RS-based RFL, which shows high degree of correlation. **a.** Simulation. **b.** Experiment.

# Conclusion

As a promising approach for investigating complex disordered system, we reported the in-depth study of the intrinsic mechanisms of the RSB in the RS-based RFL, both theoretically and experimentally for the first time. Using a $\phi$-OTDR with proprietary technologies, the RS phase fluctuations over time with unprecedented precision is measured, from which a RS-phase-variation model of RS-based RFL is proposed. Notably, from the model and the high-precision spectral measurement, we not only revealed the role of the photon phase variation in phase transition, but also observed the mode-asymmetric of the phase transition in RS-based RFL: specifically, the RSB is observed at the randomly distributed spikes (corresponding to coherent feedback-induced random longitudinal modes), and these modes display strong interactions in both the time and frequency domains; but the remaining spectral components (the slowing varying spectral profile over wavelength and time) remain in a paramagnetic state, exhibiting weak interactions.

The spin glass theory with RSB phenomenon, a challenging topic in condensed matter physics, involves complex arrangements and interactions of spin degrees of freedom. In the work, due to the high disorder of RS, RFL provides an ideal theoretical and experimental platform for the study of RSB spin glass theory. The study of this theory can not only promote the progress of condensed matter physics and deepen the understanding of the behavior of complex spin systems, but also help to resolve the quantum phase transition mechanism.

Acknowledgements

This work is supported by Natural Science Foundation of China (62075030), 111 Project (B14039) and MOST (DL2023167001L). A.S.L.G. and E.P.R. thank Brazilian funding agencies CNPq and FACEPE.

Conflict of interest

The authors declare no competing interests